\documentclass[twocolumn,preprintnumbers,prl]{revtex4}
\usepackage{amsfonts}
\usepackage[T1]{fontenc}
\usepackage{amsmath,amsbsy,amssymb,graphicx}
\usepackage{times}

\let\mathbf=\boldsymbol
\def\beginABC{\begin{subequations}}
\def\endABC{\end{subequations}}

\begin{document}

\title{{\Large Single Dirac-Cone State and Quantum Hall Effects in Honeycomb
Structure}}
\author{Motohiko Ezawa}
\affiliation{Department of Applied Physics, University of Tokyo, Hongo 7-3-1, 113-8656,
Japan }

\begin{abstract}
A honeycomb lattice system has four types of Dirac electrons corresponding
to the spin and valley degrees of freedom. We consider a state that contains
only one type of massless electrons and three types of massive ones, which
we call the single Dirac-cone state. We analyze quantum Hall (QH) effects in
this state. We make a detailed investigation of the Chern and spin-Chern
numbers. We make clear the origin of unconventional QH effects discovered in
graphene. We also show that the single Dirac-cone state may have arbitrary
large spin-Chern numbers in magnetic field. Such a state will be generated
in antiferromagnetic transition-metal oxides under electric field or
silicene with antiferromagnetic order under electric field.
\end{abstract}

\maketitle

\textit{Introduction:} Dirac electrons on a honeycomb lattice have attracted
much attention since the discovery of the unconventional quantum Hall (QH)
effect in graphene\cite{Novoselov,Kim}. Dirac electrons are ubiquitous in
monolayer honeycomb systems, where there are four types of them
corresponding to the spin and valley degrees of freedom. The spin-orbit (SO)
interaction makes Dirac electrons massive\cite{Haldane,KaneMele,LiuPRL}.
There are several materials possessing massive Dirac electrons. A remarkable
property is that we are able to control the Dirac mass externally by
applying electric field\cite{EzawaNJP}, photo-irradiation\cite%
{Oka09L,Kitagawa,EzawaPhoto} and exchange interactions\cite%
{Ryu,FengPNAS,EzawaQAHE,EzawaExM} for some cases. We can even make one type
of Dirac electrons massless and the other three massive\cite{EzawaPhoto}.
This is an intriguing state invalidating the Nielsen-Ninomiya theorem\cite%
{Nielsen}, which requires an even number of massless Dirac fermions in the
lattice system, by breaking the chiral symmetry. We have called such a state
the single Dirac-cone (SDC) state. In this work we investigate the QHE in
the SDC state and reveal some novel phenomena.

The unconventional QHE with Hall plateaux at the filling factor $\nu =\pm
2,\pm 6,\pm 10,\cdots $ implies the 4-fold degeneracy of each Landau level
in graphene. The Hall conductivity increases by $e^{2}/h$ when the Fermi
energy crosses one Landau level. If there were no degeneracy the Hall
conductivity would be half-integer quantized\cite{Zheng}, 
\begin{equation}
\sigma _{\text{H}}=\pm \frac{1}{2},\pm \frac{3}{2},\pm \frac{5}{2},\pm \frac{%
7}{2},\cdots ,  \label{AnomaSerie}
\end{equation}%
in unit of $e^{2}/h$. However the "half integer" is hidden in graphene under
the 4-fold degeneracy associated with the spin and valley degrees of freedom%
\cite{Novoselov,Kim}. There is a long history in quest for the genuine
half-integer QHE given by (\ref{AnomaSerie}). The SDC state might provide an
answer to this problem since it has only one type of massless electrons and
all nondegenerate massive electrons in general.

We start with exploring Hofstadter's butterfly diagrams\cite%
{Hatsugai93B,Hatsugai,Esaki,Sato,Hasegawa} in a SDC state. We also analyze
it in the low-energy Dirac theory, and find a good agreement between the
results in the both theories in the low magnetic field regime. We calculate
the Chern and spin-Chern numbers based on the bulk-edge correspondence\cite%
{Hatsugai93B} in the lattice theory and based on the Kubo formula\cite%
{Gusynin95L} in the Dirac theory. They show a perfect agreement in this
regime. We note that the topological insulator is indexed by the Chern and
spin-Chern numbers in external magnetic field\cite{Sheng,Prodan,Yang}, where
the time-reversal symmetry is broken but the spin $s_{z}$ is a good quantum
number. The spin-Chern number counts the Landau levels filled with the
up-spin and the down-spin electrons. 

We obtain two major findings. First, each type of electrons yields the Hall
conductivity of the form (\ref{AnomaSerie}) whether they are massless or
massive, and the total series reads $\nu =0,\pm 1,\pm 2,\pm 3,\cdots $ with
no degeneracy of each Landau level. No half-integer states appear. We also
find that the QH states may have arbitrarily high spin-Chern numbers. Note
that we only have $0,\pm 1$ in the conventional QHE. The physical reason to
allow high spin-Chern numbers is that there are only spin poralized
electrons near the Fermi level in the SDC state. A topological insulator
possessing a high spin-Chern number has never been discussed in literature.

\begin{figure}[t]
\centerline{\includegraphics[width=0.45\textwidth]{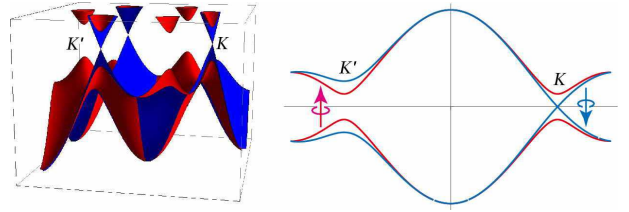}}
\caption{(a) Illustration of a SDC state. In this instance, down-spin (blue)
electrons are massless at the K point but massive at the K' point, while
up-spin (red) electrons are massive both at the K and K' points.}
\label{FigSDBand3D}
\end{figure}

\textit{Hamiltonian:} The honeycomb lattice consists of two sublattices made
of $A$ sites and $B$ sites. We consider a buckled system with the layer
separation $2\ell $ between these two sublattices. The states near the Fermi
energy are $\pi $ orbitals residing near the $K$ and $K^{\prime }$ points at
opposite corners of the hexagonal Brillouin zone. The low-energy dynamics in
the $K$ and $K^{\prime }$ valleys is described by the Dirac theory. In what
follows we use notations $s_{z}=\uparrow \downarrow $, $t_{z}=A,B$, $\eta
=K,K^{\prime }$ in indices while $s_{z}^{\alpha }=\pm 1$ for $\alpha
=\uparrow \downarrow $, $t_{z}^{i}=\pm 1$ for $i=A$,$B$, and $\eta _{i}=\pm
1 $ for $i=K,K^{\prime }$ in equations. We also use the Pauli matrices $%
\sigma _{a}$ and $\tau _{a}$ for the spin and the sublattice pseudospin,
respectively.

We investigate the system in perpendicular magnetic field $B$\ by
introducing the Peirls phase, $\Phi _{ij}=\frac{e}{h}\int_{\mathbf{r}_{i}}^{%
\mathbf{r}_{j}}\mathbf{A}\cdot d\mathbf{r}$, with $\mathbf{A}$ the magnetic
potential. Any hopping term from site $i$ to site $j$ picks up the phase
factor $e^{2\pi i\Phi _{ij}}$. The magnetic field is given by $B=2\Phi /3%
\sqrt{3}a^{2}$ in unit of $e/h$, where $a$ is the lattice constant and $\Phi 
$ is the magnetic flux penetrating one hexagonal area. Note that $\Phi =1$
implies $B=1.6\times 10^{5}$ Tesla in the case of graphene.

A generic Hamiltonian contains eight interaction terms mutually commutative
in the Dirac limit. Among them four contribute to the Dirac mass. With the
inclusion of those affecting the Dirac mass, the tight-binding model reads%
\cite{EzawaExM} 
\begin{align}
H& =-t\sum_{\left\langle i,j\right\rangle \alpha }e^{2\pi i\Phi
_{ij}}c_{i\alpha }^{\dagger }c_{j\alpha }  \notag \\
& +i\frac{\lambda _{\text{SO}}}{3\sqrt{3}}\sum_{\left\langle \!\left\langle
i,j\right\rangle \!\right\rangle \alpha }s_{z}^{\alpha }e^{2\pi i\Phi
_{ij}}\nu _{ij}c_{i\alpha }^{\dagger }c_{j\alpha }  \notag \\
& +\lambda _{V}\sum_{i\alpha }t_{z}^{i}c_{i\alpha }^{\dagger }c_{i\alpha
}+\lambda _{\text{SX}}\sum_{i\alpha }s_{z}^{\alpha }c_{i\alpha }^{\dagger
}c_{i\alpha }  \notag \\
& +i\frac{\lambda _{\text{H}}}{3\sqrt{3}}\sum_{\left\langle \!\left\langle
i,j\right\rangle \!\right\rangle \alpha }e^{2\pi i\Phi _{ij}}\nu
_{ij}c_{i\alpha }^{\dagger }c_{j\alpha },  \label{BasicHamil}
\end{align}%
where $\left\langle \!\left\langle i,j\right\rangle \!\right\rangle $ run
over all the next-nearest neighbor hopping sites. We explain each term. The
first term represents the usual nearest-neighbor hopping with the transfer
energy $t\approx 1.6$eV. The second term represents the effective SO coupling%
\cite{KaneMele} with $\lambda _{\text{SO}}\approx 3.9$meV\cite{LiuPRB},
where $\nu _{ij}=\pm 1$ if the next-nearest-neighboring hopping is
anticlockwise (clockwise) with respect to the positive $z$ axis. The third
term represents the staggered sublattice potential term with $\lambda _{V}$
due to the bucked structure\cite{KaneMele}. It may be present intrinsically
as in boron-nitride and transition metal dichalcogenides\cite%
{Xiao,Feng,FengPNAS} or generated\cite{EzawaNJP} externally by applying
external electric field $E_{z}$, where $\lambda _{V}=\ell E_{z}$. The fourth
term represents the staggered exchange term\cite{EzawaExM} with the
difference $\lambda _{\text{SX}}$ between the $A$ and $B$ sites. The fifth
term is the Haldane term\cite{Haldane}, which may be generated by photo
irradiation\cite{Oka09L,Kitagawa,EzawaPhoto}.

The low-energy Dirac Hamiltonian at the $K$ point is\cite{EzawaExM}%
\begin{eqnarray}
H_{\eta } &=&v_{\text{F}}\left( \eta P_{x}\tau _{x}+P_{y}\tau _{y}\right) 
\notag \\
&&+\lambda _{\text{SO}}\eta \tau _{z}\sigma _{z}-\lambda _{V}\tau
_{z}+\lambda _{\text{SX}}\sigma _{z}\tau _{z}+\eta \lambda _{\text{H}}\tau
_{z},  \label{TotalDirac}
\end{eqnarray}%
where $v_{\text{F}}=\frac{\sqrt{3}}{2}at$ is the Fermi velocity, and $%
P_{i}\equiv \hbar k_{i}+eA_{i}$ is the covariant momentum. When the spin $%
s_{z}$ is diagonalized, the term%
\begin{equation}
\Delta _{s_{z}}^{\eta }=\eta s_{z}\lambda _{\text{SO}}-\lambda
_{V}+s_{z}\lambda _{\text{SX}}+\eta \lambda _{\text{H}}  \label{DiracMass}
\end{equation}%
becomes the mass of Dirac electrons with $2|\Delta _{s_{z}}^{\eta }|$ being
the gap at the $K_{\eta }$ point with the spin $s_{z}$.

\textit{Single Dirac-Cone States:} We can make a full control of the Dirac
mass independently at each spin and valley. For instance, we may choose the
parameters so that%
\begin{equation}
\Delta _{\downarrow }^{K}=0\text{ with all other }\Delta _{s_{z}}^{\eta
}\neq 0,  \label{MassSDC}
\end{equation}%
which generates the SDC state as in Fig.\ref{FigSDBand3D}(a). The band
structure of a zigzag nanoribbon at $\Phi =0$ is illustrated in Fig.\ref%
{FigFanSDC}(a), where there appear only down-spin electrons near the Fermi
level at the $K$ point.

\begin{figure}[t]
\centerline{\includegraphics[width=0.5\textwidth]{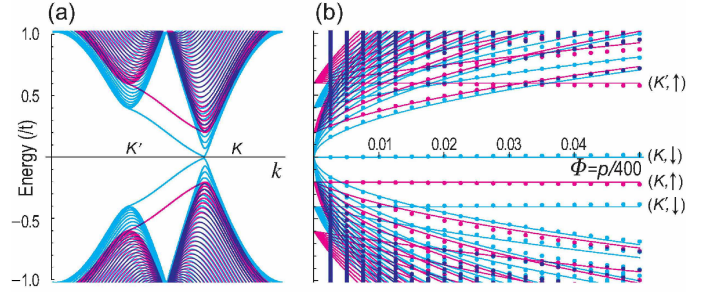}}
\caption{(a) The band structure of a zigzag nanoribbon at $\Phi =0$. There
exists only up-spin electrons near the Fermi level. We have set $\protect%
\lambda _{\text{SO}}=0.2t$ and $\protect\lambda _{\text{SX}}=-0.3t$ and $%
\protect\lambda_V=0.1t$ for illustration. The horizontal axis is the
momentum $k$. (b) A closer look of the Hofstadter's butterfly for $\Phi
=p/400$ with $p=1,2,\cdots ,20$, and the Fan diagram of Landau levels
derived in the Dirac theory. The horizontal axis is the magnetic flux $\Phi $%
.}
\label{FigFanSDC}
\end{figure}

\textit{Fan Diagram:} We introduce a pair of Landau-level ladder operators, 
\begin{equation}
\hat{a}=\frac{\ell _{B}(P_{x}+iP_{y})}{\sqrt{2}\hbar },\quad \hat{a}%
^{\dagger }=\frac{\ell _{B}(P_{x}-iP_{y})}{\sqrt{2}\hbar },  \label{G-OperaA}
\end{equation}%
satisfying $[\hat{a},\hat{a}^{\dag }]=1$, where $\ell _{B}=\sqrt{\hbar /eB}$
is the magnetic length. The Hamiltonian $H_{\eta }$ is block diagonal and
given by%
\begin{equation}
H_{\eta }=\left( 
\begin{array}{cc}
H_{\uparrow }^{\eta } & 0 \\ 
0 & H_{\downarrow }^{\eta }%
\end{array}%
\right) ,  \label{HamilBrockA}
\end{equation}%
with the diagonal elements being%
\begin{eqnarray}
H_{s_{z}}^{\text{K}} &=&\left( 
\begin{array}{cc}
\Delta _{s_{z}}^{\text{K}}\left( E_{z}\right) & \hbar \omega _{\text{c}}\hat{%
a}^{\dagger } \\ 
\hbar \omega _{\text{c}}\hat{a} & -\Delta _{s_{z}}^{\text{K}}\left(
E_{z}\right)%
\end{array}%
\right) ,  \label{BlockEleme} \\
H_{s_{z}}^{\text{K'}} &=&\left( 
\begin{array}{cc}
\Delta _{s_{z}}^{\text{K'}}\left( E_{z}\right) & -\hbar \omega _{\text{c}}%
\hat{a} \\ 
-\hbar \omega _{\text{c}}\hat{a}^{\dagger } & -\Delta _{s_{z}}^{\text{K'}%
}\left( E_{z}\right)%
\end{array}%
\right)
\end{eqnarray}%
in the basis $\left\{ \psi _{A},\psi _{B}\right\} ^{t}$.

It is straightforward to diagonalize the Hamiltonian $H_{s_{z}}^{\eta }$.
The eigenvalues are $\pm E_{s_{z}}^{\eta }(N)$ with\beginABC\label{SpectLL}%
\begin{equation}
E_{s_{z}}^{\eta }(N)=\sqrt{(\hbar \omega _{\text{c}})^{2}N+(\Delta
_{s_{z}}^{\eta })^{2}},  \label{HighLL}
\end{equation}%
for $N=1,2,\cdots $, which depend on $\Phi $. We also have%
\begin{equation}
E_{s_{z}}^{\eta }(0)=\eta \Delta _{s_{z}}^{\eta },  \label{ZeroLL}
\end{equation}%
\endABC corresponding to $N=0$, which is independent of $\Phi $: See Fig.\ref%
{FigFanSDC}(b). The eigenstate describes electrons when $E_{s_{z}}^{\eta }>0$
and holes when $E_{s_{z}}^{\eta }<0$.

We refer to each energy spectrum $\pm E_{s_{z}}^{\eta }(N)$ together with $%
E_{s_{z}}^{\eta }(0)$\ as a fan. There are four fans indexed by valley $%
K_{\eta }$ and spin $s_{z}$. Each fan consists of two parts, one for
electrons and the other for holes. These two parts are connected at one
pivot when $\Delta _{s_{z}}^{\eta }=0$, and otherwise one fan has two
pivots. The separation between these two pivots is given by $2\Delta
_{s_{z}}^{\eta }$, while the average distance of the two pivots from the
Fermi level is given by $\mu _{s_{z}}^{\eta }$. Let us call the energy level
(\ref{ZeroLL}) the lowest Landau level. In this convention there exists one
lowest Landau level in each fan. Thus there are four lowest Landau levels in
one fan diagram.

We present the fan diagram for the SDC state in Fig.\ref{FigFanSDC}(b). Four
decomposed fans are visible since the four types of Dirac electrons have
different masses $\Delta _{s_{z}}^{\eta }$ in the bulk spectrum [Fig.\ref%
{FigFanSDC}(a)]. We see also four lowest Landau levels indexed by the spin
and valley degrees of freedom ($K_{\eta },s_{z}$).

\begin{figure}[t]
\centerline{\includegraphics[width=0.5\textwidth]{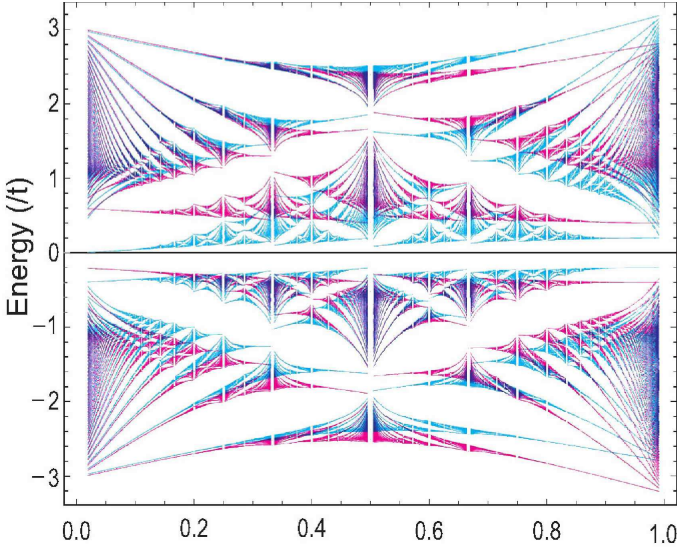}}
\caption{Spin-resolved Hofstadter's diagram in a SDC state. We have set $%
\protect\lambda _{\text{SO}}=0.2t$ and $\protect\lambda _{\text{SX}}=-0.3t$
and $\protect\lambda_V=0.1t$ for illustration. The contribution from
up(down)-spin electrons is shown in magenta (cyan). The vertical axis is the
energy in unit of $t$. The horizontal axis is the magnetic flux $\Phi $. We
have taken $\Phi =p/q$ with $q\leq 100$. }
\label{FigButterflySDC}
\end{figure}

\textit{Hofstadter Butterfly:} We compute the bulk band structure
numerically by applying periodic boundary conditions to the honeycomb
system. This requires that the magnetic flux $\Phi $ to be a rational
number, $\Phi =p/q$ ($p$ and $q$ are mutually prime integers). Then, the
system is periodic in both spatial directions. We use the Bloch theorem to
reduce the Schr\"{o}dinger equation to a $2q\times 2q$ matrix equation for
each $s_{z}=\uparrow \downarrow $, where the factor $2$ is due to the
sublattice ($A$,$B$) degrees of freedom. In so doing we choose a generalized
gauge of the one used in graphene\cite{Hatsugai93B} so as to include the
link connecting the next-nearest neighbor hopping sites. It is given in such
a way that the magnetic flux becomes $1/6$ for each isosceles triangle whose
two edges are given by the neighbor hopping.

The resulting band structure is the Hofstadters butterfly diagram, which we
display for the SDC state in Fig.\ref{FigButterflySDC}. We present a closer
look of the Hofstadter butterfly in the low magnetic field regime ($\Phi
<5/100$) in Fig.\ref{FigFanSDC}(b) together with the fan diagram. The
spectra implied by the Hofstadter butterfly and the fan diagram agree one to
another quite well for $\Phi <1/100$. The agreement is very good for the
lowest and first Landau levels for a wide range of $\Phi $.

\textit{Topological Charges and Conductance: }The Hall and spin-Hall
conductivities are given by using the TKNN formula\cite{TKNN}, 
\begin{equation}
\sigma _{xy}=\frac{e^{2}}{2\pi \hbar }\mathcal{C},\qquad \sigma _{xy}^{\text{%
spin}}=\frac{e}{2\pi \hbar }\mathcal{C}_{\text{spin}},
\end{equation}%
where $\mathcal{C}$ and $\mathcal{C}_{\text{spin}}$\ are the Chern number
and the spin-Chern number, respectively. A topological insulator is indexed
by a set of these two topological charges. When spin $s_{z}$ is a good
quantum number, they are given by\beginABC\label{ChernS}%
\begin{eqnarray}
\mathcal{C} &=&\mathcal{C}_{\uparrow }^{K}+\mathcal{C}_{\uparrow
}^{K^{\prime }}+\mathcal{C}_{\downarrow }^{K}+\mathcal{C}_{\downarrow
}^{K^{\prime }}, \\
\mathcal{C}_{\text{spin}} &=&\frac{1}{2}(\mathcal{C}_{\uparrow }^{K}+%
\mathcal{C}_{\uparrow }^{K^{\prime }}-\mathcal{C}_{\downarrow }^{K}-\mathcal{%
C}_{\downarrow }^{K^{\prime }}),
\end{eqnarray}%
\endABC where $\mathcal{C}_{s_{z}}^{\eta }$ is the summation of the Berry
curvature in the momentum space over all occupied states of electrons with
spin $s_{z}$ in the $K_{\eta }$ valley.

\begin{figure}[t]
\centerline{\includegraphics[width=0.5\textwidth]{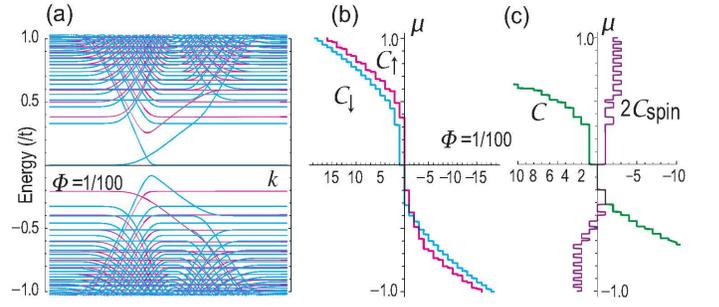}}
\caption{(a) The energy spectra of the honeycomb lattice with zigzag edges
for $\Phi =1/100$. The horizontal axis is the momentum $k$. We can determine
the topological number $\mathcal{C}_{s_{z}}$ by examining the edge modes.
(b) The topological number $\mathcal{C}_{s_{z}}$ calculated based on the
Kubo formula (\protect\ref{DiracKubo}) as a function of $\protect\mu $. The
horizontal axis is the topological number $\mathcal{C}_{s}$. The results
based on the bulk-edge correspondence (a) and the Kubo formula (b) are
identical. (c) The Chern and spin-Chern numbers $\mathcal{C}$ (green) and $2%
\mathcal{C}_{\text{spin}}$ (violet) derived based on (\protect\ref{ChernS}). 
}
\label{FigChernSCD}
\end{figure}

The most convenient way to determine the topological charge in the lattice
formulation is to employ the bulk-edge correspondence\cite{Hatsugai93B}. The
edge-state analysis can be performed for a system with boundaries such as a
cylinder. When solving the Harper equation on a cylinder, the spectrum
consists of bulk bands and topological edge states as in Fig.\ref%
{FigChernSCD}(a). We typically find a few edge states within the bulk gaps,
some of which cross the gap from one bulk band to another. Each edge state
contributes one unit to the quantum number $\mathcal{C}_{s_{z}}$ for each $%
s_{z}=\uparrow \downarrow $. More precisely, in order to evaluate $\mathcal{C%
}_{s_{z}}$, we count the edge states, taking into account their location
(right or left edges) and direction (up or down) of propagation\cite%
{Hatsugai93B}. The location of each state is derived by computing the wave
function, while the direction of propagation can be obtained from the sign
of its momentum derivative $dE/dk$, with $k$ the momentum parallel to the
edge.

We focus on one edge. Edge states with opposite directions contribute with
opposite signs. The resultant formula reads%
\begin{equation}
\mathcal{C}_{s_{z}}=N_{\text{u}}^{s_{z}}-N_{\text{d}}^{s_{z}},
\label{BerryFormu}
\end{equation}%
where $N_{\text{u}}^{s_{z}}$ and $N_{\text{d}}^{s_{z}}$ denote the number of
up- and down-moving states with spin $s_{z}$, respectively, at the right
edge.

It is also possible to use the Kubo formulation in the Dirac theory to
derive the Hall conductivity for each spin $s_{z}$ in each valley $K_{\eta }$%
. Such a formula has been derived for graphene\cite{Gusynin95L}. We may
generalize it and apply it to the Dirac system (\ref{TotalDirac}),%
\begin{align}
\mathcal{C}_{s_{z}}^{\eta }(\mu )=& \frac{1}{4}\Big[\tanh \frac{\mu +\Delta
_{s_{z}}^{\eta }}{2k_{\text{B}}T}+\tanh \frac{\mu -\Delta _{s_{z}}^{\eta }}{%
2k_{\text{B}}T}\Big]  \notag \\
& +\frac{1}{2}\sum_{N=1}^{\infty }\Big[\tanh \frac{\mu +E_{s_{z}}^{\eta }(N)%
}{2k_{\text{B}}T}+\tanh \frac{\mu -E_{s_{z}}^{\eta }(N)}{2k_{\text{B}}T}\Big]%
,  \label{DiracKubo}
\end{align}%
where $\mu $ is the chemical potential.

As a clear illustration we present the result of the edge-state analysis in
magnetic field at $\Phi =1/100$ in Fig.\ref{FigChernSCD}(a). According to
the formula (\ref{BerryFormu}) we count the number of edge modes, from which
we derive the topological numbers $\mathcal{C}_{\uparrow }^{N}$ (magenta)
and $\mathcal{C}_{\downarrow }^{N}$\ (cyan) based on (\ref{ChernS}). On the
other hand we may calculate them from the Kubo formula (\ref{DiracKubo}),
which we give in Fig.\ref{FigChernSCD}(b). We can explicitly check that they
agree one to another. The Chern and spin-Chern numbers $\mathcal{C}^{N}$
(blue) and $\mathcal{C}_{\text{spin}}^{N}$\ (green) are calculated from $%
\mathcal{C}_{\uparrow }^{N}$ and $\mathcal{C}_{\downarrow }^{N}$\ in Fig.\ref%
{FigChernSCD}(c).

\begin{figure}[t]
\centerline{\includegraphics[width=0.34\textwidth]{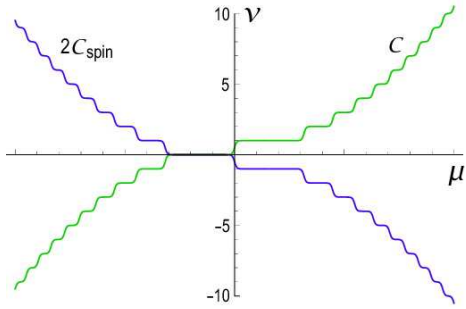}}
\caption{The Chern and spin-Chern numbers $\mathcal{C}$ (green) and $2%
\mathcal{C}_{\text{spin}}$ (violet) of the QH state for $B\lesssim 16$Tesla as a
function of the chemical potential $\protect\mu .$}
\label{FigChernKK}
\end{figure}

The experimentally accessible regime is the low magnetic field limit $\Phi
\lesssim 10^{-4}$ or $B\lesssim 16$Tesla. The prominent feature is that only
the down-spin fan is present near the Fermi level both in the electron and
hole sectors as in Fig.\ref{FigFanSDC}(b). Consequently all the QH states
are made of down-spin electrons near the Fermi level. They contribute
equally to the Chern and spin-Chern numbers: It follows from (\ref{ChernS})
that $\mathcal{C}=\mathcal{C}_{\downarrow }^{K}=-2\mathcal{C}_{\text{spin}}$%
. As shown in Fig.\ref{FigChernKK}, the series of QH plateaux reads $%
\nu=0,\pm 1,\pm 2,\pm 3,\pm 4,\cdots $ with no degeneracy in each level,
where the spin-Chern number reads%
\begin{equation}
\mathcal{C}_{\text{spin}}=-\frac{1}{2}\nu .
\end{equation}%
The maximum value of $|\mathcal{C}_{\text{spin}}|$ increases as $\Phi $
becomes lower.

\textit{Discussions:} Our analysis on the SDC state is applicable to any
Dirac systems described by the Hamiltonian (\ref{BasicHamil}) or (\ref%
{TotalDirac}). Here we address the problem how to materialize a SDC state.
Transition-metal oxide grown on [111] direction would be the first candidate%
\cite{Hu}. A salient property is that the material contains an intrinsic
staggered exchange effect $\varpropto \lambda _{\text{SX}}$. It has
antiferromagnetic order yielding a Dirac mass. We can control the band
structure by applying electric field due to the buckled structure.

We may also consider silicene with antiferromagnetic order ($\lambda _{\text{%
SX}}\neq 0$) introduced by a proximity coupling effect method\cite{EzawaExM}%
. Alternatively we may apply photo irradiation\cite{EzawaPhoto} to produce
the Haldane term ($\lambda _{\text{H}}\neq 0$). In so doing it is necessary
to arrange that transitions between Landau levels within the conduction band
as well as between the valence and conduction bands (interband transitions)
are prohibitted. This would be possible if we use photo-irradiation with
frequency $\omega $ such that $\hbar \omega >3$eV since the Landau levels
are bounded within this range as in Fig.\ref{FigButterflySDC}.

\textit{Conclusions:} We have analyzed the QH effects in the SDC state. We
conclude that no half-integer QH states appear because there are four types
of electrons each of which contributes the basic series (\ref{AnomaSerie})
whether they are massless or massive. This must be the case in any lattice
theory since the number of electron types is even whether they are massless
or massive. We have also found that the SDC state has arbitrarily high
spin-Chern numbers in magnetic field. 

I am very much grateful to N. Nagaosa, H. Aoki and Y. Hatsugai for many
fruitful discussions on the subject. This work was supported in part by
Grants-in-Aid for Scientific Research from the Ministry of Education,
Science, Sports and Culture No. 22740196.

\end{document}